\documentclass{article}
\usepackage{amsmath}[1996/11/01]
\usepackage{amssymb,amsthm,amsxtra}
\usepackage{epsfig}

\newlength{\mytopmargin}
\newlength{\myleftmargin}
\def\zz{\rlx\hbox{\small \sf Z\kern-.4em Z}}

\newtheorem{lemma}{Lemma}[section]

\newtheorem{prop}[lemma]{Proposition}

\begin{document}

\vspace{1cm}
\noindent
\begin{center}{   \large \bf Eigenvalue distributions for some correlated complex sample
covariance matrices} 
\end{center}
\vspace{5mm}

\noindent
\begin{center}
 P.J.~Forrester\\

\it Department of Mathematics and Statistics, \\
University of Melbourne, Victoria
3010, Australia
\end{center}
\vspace{.5cm}
\begin{quote}
The distributions of the smallest and largest eigenvalues for the matrix product
$Z^\dagger Z$, where $Z$ is an $n \times m$ complex Gaussian matrix with correlations
both along rows and down columns, are expressed as $m \times m$ determinants. In the
case of correlation along rows, these expressions are computationally more efficient
than those involving sums over partitions and Schur polynomials reported recently
for the same distributions.
\end{quote}

\vspace{.5cm}
\noindent
\section{Introduction}
\setcounter{equation}{0}
A typical setting in multivariate statistics is to measure each of $m$ variables
$x_1,\dots,x_m$ a total of $N$ times. For example, the variable $x_k$ may denote the
wind speed at weather station $k$ at a specific time of day; recording the value on
successive days gives a sequence of values $x_k^{(j)}$, $j=1,\dots,N$ for the variable
$x_k$, which forms a column vector $\vec{x}_k = [x_k^{(j)}]_{j=1,\dots,N}$.
Collecting together the column vectors for each of the variables $x_k$ gives
the data matrix $X = [\vec{x}_k]_{k=1,\dots,m}$. Let the average of the readings of
variable $x_k$ be denoted $\bar{x}_k$, so that
$$
\bar{x}_k = {1 \over N} \sum_{j=1}^N x_j.
$$
Let $\vec{\bar{x}}_k = [\bar{x}_k]_{j=1,\dots,N}$ be the corresponding (constant) column
vector, and set $\bar{X} := [\vec{\bar{x}}_k]_{k=1,\dots,m}$. Forming now
$$
{1 \over n} A := {1 \over n} (X - \bar{X})^T (X - \bar{X}) =
\Big [ {1 \over n} \sum_{j=1}^N (x_{k_1}^{(j)} - \bar{x}_{k_1}^{(j)})
(x_{k_2}^{(j)} - \bar{x}_{k_2}^{(j)}) \Big ]_{k_1,k_2 = 1, \dots, m},
$$
$n= N - 1$, gives an empirical approximation to the covariance matrix
$ [ \langle (x_{k_1} - \langle {x}_{k_1} \rangle)
 (x_{k_2} - \langle {x}_{k_2} \rangle) ]_{k_1,k_2=1,\dots,m}$ for the variables
$\{x_k\}_{k=1,\dots,m}$.

Analytic studies of the matrix $A$ can be carried out in the case that the variables
$x_1,\dots,x_m$ relating to the data matrix $X$ are chosen from a multivariate Gaussian
distribution with variance matrix $\Sigma$ and mean $\vec{\mu}$. Then it is known
(see e.g.~\cite{GN99}) that the distribution of $A$ is the same as that for the matrix
product $Y^T Y$, where $Y$ is an $m \times n$, $n = N-1$, Gaussian matrix in which each
row is drawn from a multivariate Gaussian distribution with covariance matrix $\Sigma$
and mean zero. Thus the joint probability density function (p.d.f.) of the elements of
$Y$ is
\begin{equation}\label{YY}
{1 \over C} e^{-{\rm Tr} ( \Sigma^{-1}Y^T  Y/2)},
\end{equation}
where here and throughout (unless otherwise stated) $C$ represents {\it some}
constant (i.e.~quantity independent of the main variables of the equation, which
here are the elements of $Y$).

A standard practice in studying the empirical covariance matrix is to form the
eigenvalue-eigenvector decomposition. This comes under the name of principal component
analysis (see e.g.~\cite{Jo01}). On a theoretical front one seeks analytic forms for
eigenvalue distributions of the matrix $A = Y^T Y$ when $Y$ is distributed according to
(\ref{YY}). In fact the eigenvalue p.d.f.~can be written down in terms of a 
multivariable generalized
hypergeometric function based on zonal polynomials (see e.g.~\cite{Mu82}). This
function is inherently difficult to compute, but there have been some recent advances
\cite{KE04}. It is also possible to integrate over this p.d.f.~to express the distribution
of the largest eigenvalue as another generalized hypergeometric function 
\cite{Ja64}.

In a recent work \cite{RVA05} a study of the p.d.f.~for the smallest and largest
eigenvalues of the matrix $A = Z^\dagger Z$ for
$Z$ an $n \times m$, $(n \ge m)$ complex Gaussian matrix with p.d.f.~
\begin{equation}\label{ZZ}
{1 \over C} e^{-{\rm Tr} ( \Sigma^{-1} Z^\dagger Z)}
\end{equation}
has been undertaken.
Earlier studies had considered these p.d.f.'s in the case $\Sigma = I$
\cite{Fo93c,FH94}.
The setting of complex data matrices
is of great importance in recent quantitative studies of wireless
communication (see e.g.~\cite{TV04,SMM05}). A significant feature of the p.d.f.~(\ref{ZZ}) is
that the corresponding joint eigenvalue p.d.f.~of $A$ can be written as a determinant
\cite{GS00,BBP05,BK04,SMM05}. Moreover, as to be shown in Section 2 below,
the marginal distributions by way of the p.d.f.~of the smallest
and largest eigenvalues can also be evaluated as determinants. In contrast, these same
distributions where evaluated in  \cite{RVA05} as a sum over Schur polynomials
and as a generalized hypergeometric function based on the Schur polynomials respectively
(see (\ref{Es}) and (\ref{f2}) below). 

Suppose more generally that the complex data matrix $Z$ has p.d.f.
\begin{equation}\label{ZZ1}
{1 \over C} e^{-{\rm Tr} ( \Sigma_1^{-1} Z^\dagger \Sigma_2^{-1} Z)}.
\end{equation}
Here $\Sigma_2$ can be interpreted as the covariance coupling the measurements of a single
variable $z_k$. Very recently \cite{SM04,SMM05}, it has been shown that for this distribution
the canonical average
\begin{equation}\label{can}
\Big \langle \det (1 + u Z^\dagger Z)^p \Big \rangle
\end{equation}
can be expressed as an $n \times n$ determinant, even though the joint eigenvalue p.d.f.~of
(\ref{ZZ}) cannot itself be written in a determinant form. In Section 3 we will use the
method of \cite{SMM05} to similarly express the p.d.f.~for the smallest and largest
eigenvalues of $Z^\dagger Z$ with $Z$ distibuted as (\ref{ZZ1}) in terms of determinants.

\section{Case of a single covariance matrix}
\setcounter{equation}{0}
\subsection{Correlation across rows of $Z$} 
Consider the p.d.f.~(\ref{ZZ}). Introduce the singular value decomposition
\begin{equation}\label{ZUV}
Z = U {\rm diag}(\mu_1,\dots, \mu_m) V,
\end{equation}
where $U$ ($V$) is a $m \times m$ ($n \times n$) unitary matrix and the
$\mu_j^2 =: \lambda_j$ are the eigenvalues of the positive definite matrix $Z^\dagger Z$.

We seek the joint distribution of the $\{\lambda_j\}_{j=1,\dots,m}$,
$p(\lambda_1,\dots,\lambda_m)$ say. Firstly, with $A = Z^\dagger Z$, we know
(see e.g.~\cite{Fo02})
\begin{eqnarray}\label{5e}
(d A) & = & {1 \over C} \det (Z^\dagger Z)^{n-m} (d Z) \nonumber \\
 & = &  {1 \over C} \prod_{j=1}^m \lambda_j^{n-m} \prod_{1 \le j < k \le m}
(\lambda_k - \lambda_j)^2 d \lambda_1 \cdots d \lambda_m (V^\dagger d V),
\end{eqnarray}
where $(V^\dagger dV)$ is the Haar measure (uniform distribution) on the space of
$m \times m$ unitary matrices $U(m)$. Thus
\begin{equation}\label{5f}
p(\lambda_1,\dots,\lambda_m) 
 = {1 \over C}
\prod_{j=1}^m \lambda_j^{n-m} \prod_{1 \le j < k \le m}
(\lambda_k - \lambda_j)^2  \int_{V \in U(m)}
e^{-{\rm Tr} (\Sigma^{-1} V^\dagger {\rm diag}(\lambda_1,\dots,\lambda_m) V)}
(V^\dagger dV).
\end{equation}
This is the well known Harish-Chandra/Itzykson-Zuber matrix integral (see e.g.~\cite{Or04}).
It has a closed form determinantal evaluation, which when substituted in (\ref{5f}) implies
\begin{equation}\label{6.1}
 p(\lambda_1,\dots,\lambda_m) = {1 \over C} \prod_{j=1}^m \lambda_j^{n-m} \prod_{1 \le j < k \le m}
{ (\lambda_k - \lambda_j) \over (s_k - s_j) }
\det [ e^{-s_j \lambda_k} ]_{j,k=1,\dots,m}
\end{equation}
where $\{s_1,\dots,s_m\}$ are the eigenvalues of $\Sigma^{-1}$. 
As referenced in the third sentence below (\ref{ZZ}), the 
result (\ref{6.1}) has been made explicit in a number of recent works.

Consider now the probability $E((\lambda,\infty))$ that the interval $(\lambda,\infty)$ is
free of eigenvalues. This is related to the p.d.f.~of the largest eigenvalue,
 $p^{\rm max}(\lambda)$
say, by
$$
p^{\rm max}(\lambda) = - {d \over d \lambda} E((\lambda,\infty)).
$$
We have
\begin{eqnarray*}
E((\lambda,\infty)) & := & 
\int_0^\lambda d\lambda_1 \cdots \int_0^\lambda d\lambda_m \,
p(\lambda_1,\dots,\lambda_m) \nonumber \\
& = & {1 \over C} {1 \over \prod_{j < k}^m (s_k - s_j) }
\int_0^\lambda d\lambda_1 \cdots \int_0^\lambda d\lambda_m \,
\prod_{j=1}^m \lambda_j^{n-m}  \prod_{j < k}^m (\lambda_k - \lambda_j)
\det [ e^{-s_j \lambda_k} ]_{j,k=1,\dots,m}.
\end{eqnarray*}
Because both factors in the integrand are anti-symmetric in $\{\lambda_j\}_{j=1,\dots,m}$, and
$$
\prod_{j < k}^m (\lambda_k - \lambda_j) = {\rm Asym} \, \lambda_1^0 \lambda_2 \cdots
\lambda_m^{m-1},
$$
the product can be replaced by $ \lambda_1^0 \lambda_2 \cdots
\lambda_m^{m-1}$ provided we multiply by $m!$. Doing this we see the integrations over
$\{\lambda_k\}$ can be performed column by column, to give
\begin{equation}\label{7.1}
E((\lambda,\infty)) = {m! \over C}
 {1 \over \prod_{j < k}^m (s_k - s_j) }
\det \Big [ \int_0^\lambda t^{n-m+k-1} e^{-s_j t} \, dt \Big ]_{j,k=1,\dots,m}.
\end{equation}
To evaluate $C$, we note $\lim_{\lambda \to \infty} E((\lambda,\infty)) = 1$.
The integral in (\ref{7.1}) can be evaluated in this limit to give
$$
1 =  {m! \over C}
 {1 \over \prod_{j < k}^m (s_k - s_j) }
\det \Big [ s_j^{-(n-m+k)} (n-m+k-1)! \Big ]_{j,k=1,\dots,m}.
$$
Factoring the factorials from the determinant and then making use of the
Vandermonde determinant formula shows
\begin{equation}\label{C}
C = (-1)^{m(m-1)/2} m! \prod_{k=1}^m (n-m+k-1)! \prod_{j=1}^m s_j^{-n}.
\end{equation}
Substituting this in (\ref{7.1}), and changing variables $s_j \mapsto \lambda s_j$ in the
integral therein, we obtain for our final expression
\begin{equation}\label{7.1a}
E((\lambda,\infty)) = {1 \over \prod_{k=1}^m (n - m + k -1)! } 
{ \prod_{j=1}^m (\lambda s_j)^{n} \over \prod_{j < k}^m (-\lambda )(s_k - s_j) }
\det  \Big [ \int_0^1 t^{n-m+k-1} e^{-\lambda s_j t} \, dt \Big ]_{j,k=1,\dots,m}. 
\end{equation}

We remark that in the case that $s_j=1$ $(j=r+1,\dots,m)$, the $m \to \infty$ limit of
$E((\lambda,\infty))$, with $\lambda, s_1,\dots, s_r$ appropriately scaled, is studied in
\cite{BBP05}. We remark too that in \cite[Corollary 3.3]{RVA05}
$E((\lambda,\infty))$ is expressed in terms of the generalized multi-variable
hypergeometric function 
\begin{equation}\label{f11}
{}_1 F_1(a,b;x_1,\dots,x_m) := \sum_{\kappa} {[a]_{\kappa} \over d_\kappa' [b]_\kappa}
s_\kappa(x_1,\dots,x_m).
\end{equation}
In (\ref{f11}) $s_\kappa$ denotes the Schur polynomial labelled by a partition
$\kappa = (\kappa_1,\dots,\kappa_m)$, $\kappa_1 \ge \dots \ge \kappa_m$,
$$
[a]_\kappa := \prod_{j=1}^m {\Gamma(a - j+1 + \kappa_j) \over \Gamma(a - j + 1) },
$$
while
$$
d_\kappa' = {[m]_\kappa \over \bar{f}_m(\kappa)}, \qquad
 \bar{f}_m(\kappa) := \prod_{1 \le i < j \le m}
{(j-i + \kappa_i - \kappa_j )\over j - i}.
$$
Thus from \cite[Eq.~(3.5)]{RVA05}
\begin{equation}\label{f2}
E((\lambda,\infty)) = \prod_{k=1}^m {\Gamma(k) \over
\Gamma(n+k) }
 \prod_{j=1}^m (\lambda s_j)^n \, {}_1 F_1(n;n+m;-\lambda s_1,\dots, - \lambda s_m).
\end{equation}
Comparing (\ref{f2}) and (\ref{7.1}) gives the determinant formula 
\begin{equation}\label{f3}
 F_1(n;n+m;x_1,\dots,  x_m) =
\prod_{k=1}^m {\Gamma(n+k) \over \Gamma(k) \Gamma(n-m+k) }
{1 \over \prod_{j<k}^m (x_k - x_j) }
\det  \Big [ \int_0^1 t^{n-m+k-1} e^{x_j t} \, dt \Big ]_{j,k=1,\dots,m}.
\end{equation}
The integral in (\ref{7.1a}) is itself a special case of a one variable confluent
hypergeometric function ${}_1 F_1$, allowing us to write
\begin{eqnarray}\label{2.10a}
E((\lambda, \infty)) & = & {1 \over \prod_{k=1}^m (n-m+k)! }
{\prod_{j=1}^m (\lambda s_j)^n \over \prod_{j<k}^m (-\lambda)(s_k - s_j) }
\nonumber \\
&& \qquad \times
\det \Big [ {}_1 F_1(n-m+k;n-m+k+1;-\lambda s_j ) \Big ]_{j,k=1,\dots,m}.
\end{eqnarray}
Note that (\ref{2.10a}) and (\ref{f2}) are identical in the case $m=1$. 

We draw attention to a limiting feature of (\ref{7.1}) which is of relevance in
the study of $E((\lambda,\infty))$ for fully correlated matrices (\ref{ZZ1}).
Suppose then that $n=m$ in (\ref{7.1}), and consider the limit $s_n \to \infty$.
Integrating the final row of integrals by parts, we see that the dominant term is
that in the first column. Expanding by this term shows
$$
\lim_{s_n \to \infty} E((\lambda,\infty)) \Big |_{m = n} =
 E((\lambda,\infty)) \Big |_{m = n - 1},
$$
and iterating this we have
$$
\lim_{s_{n-m}, \dots, s_n \to \infty} E((\lambda,\infty)) \Big |_{m = n} =
 E((\lambda,\infty))
$$
where on the right hand side $E((\lambda,\infty))$ is for the general $m$
case, as given by (\ref{7.1a}).
The understanding of this result is that with $s_{n-m}, \dots, s_n \to \infty$,
the final $n-m$
rows of $Z$ become zero and so the eigenvalues of $Z^\dagger Z$ are those of
$W^\dagger W$ for $W$ the restriction of $Z$ to its first $m$ rows, together with
$m$ zero eigenvalues.

For the probability $E((0,\lambda))$ that the interval $(0,\lambda)$ is free of
eigenvalues, related to the p.d.f.~of the smallest eigenvalue, $p^{\rm min}(\lambda)$ say,
by
$$
p^{\rm min}(\lambda) = {d \over d \lambda} E((0,\lambda)),
$$
we have
\begin{eqnarray*}
E((0,\lambda)) & := & \int_\lambda^\infty d \lambda_1 \cdots
\int_\lambda^\infty d \lambda_m \, p( \lambda_1, \dots, \lambda_m)
\nonumber \\
& = & {1 \over C}
{1 \over \prod_{j < k} (s_k - s_j) }
\int_\lambda^\infty d \lambda_1 \cdots  d \lambda_m \,
\prod_{j=1}^m \lambda_j^{n-m} \prod_{1 \le j < k \le m} (\lambda_k - \lambda_j)
\det [ e^{- s_j \lambda_k} ]_{j,k=1,\dots,m}.
\end{eqnarray*}
Proceeding now as in the derivation of (\ref{7.1a}) shows
\begin{eqnarray}\label{9.1}
 E((0,\lambda)) & = & {m! \over C} { 1
 \over  \prod_{j < k}^m (s_k - s_j) } 
\det \Big [ \int_\lambda^\infty t^{n-m+k-1} e^{-s_j t} \, dt
\Big ]_{j,k=1,\dots,m} \nonumber \\
 & = & {m! \over C} {e^{-\lambda \sum_{j=1}^m s_j}
 \over  \prod_{j < k}^m (s_k - s_j) }
\det \Big [ \int_0^\infty (t+\lambda)^{n-m+k-1} e^{-s_j t} \, dt
\Big ]_{j,k=1,\dots,m}
\end{eqnarray}
where $C$ is given by (\ref{C}).
Note that for $m=n$ the determinant only contributes a constant (i.e.~term independent of
$\lambda$) and we have
\begin{equation}\label{2.12}
 E((0,\lambda)) \Big |_{m = n} = e^{- \lambda \sum_{j=1}^n s_j},
\end{equation}
which generalizes the same result known for $s_1= \cdots =s_n = 1$
\cite{Ed88, Fo93c}.

In terms of the confluent hypergeometric function
$$
U(a,b,z):= {1 \over \Gamma(a)} \int_0^\infty e^{-zt} t^{a-1}(1+t)^{b-a-1} \, dt,
$$
which unlike ${}_1 F_1(a,b;z)$ is singular at $z=0$, but like ${}_1 F_1$ is available
as an inbuilt function on a number of mathematical computing packages, (\ref{9.1})
reads
\begin{eqnarray}
 E((0,\lambda)) & = & (-1)^{m(m-1)/2} \prod_{k=1}^m {(\lambda s_k)^n \over 
\Gamma(n-m+k) } 
 {e^{-\lambda \sum_{j=1}^m s_j}
 \over  \prod_{j < k}^m (s_k - s_j) } \nonumber \\
&& \times \det \Big [ U(1,n-m+k+1,\lambda s_j) ]_{j,k=1,\dots,m}.
\end{eqnarray}
The computationally more complex evaluation
\begin{equation}\label{Es}
 E((0,\lambda)) = e^{- \lambda \sum_{j=1}^m s_j }
\sum_{k=0}^{m(n-m)} \lambda^k \sum_{\kappa: |\kappa| = k \atop \kappa_1 \le n - m}
{s_\kappa(s_1,\dots,s_m) \over d_\kappa'}
\end{equation}
is given in \cite[Eq.~(3.8)]{RVA05}.

Also given in \cite{RVA05} is an expression involving sums over partitions and
Schur polynomials for the probability
$E((0,a) \cup (b,\infty))$ that there are no eigenvalues in either of the intervals
$(0,a)$ or $(b,\infty)$. In terms of this quantity the joint p.d.f.~for the smallest
and largest eigenvalues, $p(a,b)$ say, is given by
$$
p(a,b) = - {\partial^2 \over \partial a  \partial b} E((0,a) \cup (b,\infty)).
$$
From $p(a,b)$ one can deduce the distribution of $b/a$, which is the square of the
condition number and so is of relevance in numerical analysis. The method of derivation
of (\ref{7.1a}) yields the determinant evaluation
\begin{eqnarray}
&&E((0,a) \cup (b,\infty)) \nonumber \\
&  &  \quad = {1 \over C}
{1 \over  \prod_{j < k}^m (s_k - s_j) } \int_a^b d\lambda_1 \cdots
 \int_a^b d\lambda_m \, \prod_{j=1}^m \lambda_j^{n-m} \,
\prod_{j<k}^m (\lambda_k - \lambda_j) \det [ e^{- s_j \lambda_k} ]_{j,k=1,\dots,m}
\nonumber \\
 && \quad =  {m! \over C}
{1 \over  \prod_{j < k}^m (s_k - s_j) }
\det \Big [ \int_a^b t^{n-m+k-1} e^{-s_j t} \, dt \Big ]_{j,k=1,\dots,m}
\end{eqnarray}
where $C$ is given by (\ref{C}).

\subsection{Correlation down columns of $Z$} 
Consider next the case that $Z$ has distribution
\begin{equation}\label{10.1}
{1 \over C} e^{- {\rm Tr} (Z^\dagger \Sigma_2^{-1} Z)}
\end{equation}
((\ref{ZZ1}) with $\Sigma_1 = I$). Now
\begin{equation}\label{10.2}
 e^{- {\rm Tr} (Z^\dagger \Sigma_2^{-1} Z)} =
 e^{- {\rm Tr} (  \Sigma_2^{-1} Z Z^\dagger)}
\end{equation}
and the non-zero eigenvalues of $Z Z^\dagger$ agree with the eigenvalues of $Z^\dagger Z$,
which we again denote $\{\lambda_j \}_{j=1,\dots,m}$. With $A = Z Z^\dagger$ (\ref{5e})
again applies but with $V$ replaced by $U$ (recall (\ref{ZUV})), and thus
\begin{eqnarray}\label{10.3}
&& p(\lambda_1,\dots, \lambda_m)  \nonumber \\
&& \qquad 
= {1 \over C} \prod_{j=1}^m \lambda_j^{n-m}
\prod_{1 \le j < k \le m} (\lambda_k - \lambda_j)^2 
\lim_{\lambda_{m+1}, \dots, \lambda_n \to 0}
\int_{U \in U(n) }
e^{- {\rm Tr} ( \Sigma_2^{-1} U {\rm diag}(\lambda_1,\dots, \lambda_n) U^\dagger) }
(U^\dagger dU). \nonumber \\
\end{eqnarray}
(Here the limit could have been taken immediately, but as noted in \cite{SMM05} there
are computational advantages in delaying this step.) Proceeding as in the derivation of
(\ref{6.1}) gives
\begin{eqnarray*}
&& p(\lambda_1,\dots, \lambda_m) = {1 \over C} \lim_{\lambda_{m+1}, \dots, \lambda_n \to 0}
{ \prod_{j=1}^m \lambda_j^{n-m} \prod_{j < k}^m (\lambda_k - \lambda_j)^2 \over
\prod_{j < k}^n (\lambda_k - \lambda_j) (s_k - s_j) }
\det [ e^{-s_j \lambda_k} ]_{j,k=1,\dots,n}
\end{eqnarray*}
where here $\{s_1,\dots, s_n\}$ denotes the eigenvalues of $\Sigma_2^{-1}$. 
Now taking the limit
this reads
\begin{equation}\label{11}
 p(\lambda_1,\dots, \lambda_m) = {1 \over C}
{ \prod_{j<k}^m (\lambda_k - \lambda_j) \over
\prod_{j<k}^n (s_k - s_j) } 
\det \Big [  [e^{-s_j \lambda_k} ]_{j=1,\dots,n \atop k=1,\dots,m} \quad
[s_j^{k-1}  ]_{j=1,\dots,n \atop k=1,\dots,n-m} \Big ].
\end{equation}
The joint p.d.f.~(\ref{11}) has been derived previously in \cite{CWZ03,SMM05}.

The derivations of (\ref{7.1a}) and (\ref{9.1}) can be applied to (\ref{11}) to
deduce determinant formulas for $E((\lambda,\infty))$ and $E((0,\lambda))$. Thus we find
\begin{equation}\label{11a}
E((\lambda, \infty)) = {m! \over C}
{ 1 \over
\prod_{j<k}^n (s_k - s_j) } 
\det \Big [  \Big [
\int_0^\lambda t^{k-1} e^{-s_j t} \, dt \Big ]_{j=1,\dots,n \atop k=1,\dots,m} \: \:
[s_j^{k-1} ]_{j=1,\dots,n \atop k=1,\dots,n-m} \Big ]
\end{equation}
and
\begin{equation}\label{11b}
E((0,\lambda)) = {\prod_{k=1}^m k! \over C}
{ e^{-\lambda \sum_{j=1}^n s_j } \over
\prod_{j<k}^n (s_k - s_j) } 
\det \Big [ [s_j^{-k}]_{j=1,\dots,n \atop k =1,\dots,m} \quad 
[e^{\lambda s_j} s_j^{k-1} ]_{j=1,\dots,n \atop k=1,\dots,n-m} \Big ]
\end{equation}
where
\begin{equation}\label{2.18a}
C = \prod_{k=1}^m k! \prod_{j=1}^n s_j^{-m}. 
\end{equation}

\section{Fully correlated case}
\setcounter{equation}{0}
We turn our attention now to the case (\ref{ZZ1}), in which the data matrix
$Z$ is correlated both across rows and down columns. 
Here it does not appear possible to write the joint eigenvalue p.d.f.~of 
$Z^\dagger Z$ in determinant form. 
Nonetheless, it has been shown
recently by Simon and Moustakos \cite{SM04} (see \cite{SMM05} for a detailed presentation)
that it is possible to give a determinant formula for the average (\ref{can}). Here we
will show that their calculation can be adopted to give determinant formulas for
$E((\lambda,\infty))$ and $E((0,\lambda))$ (the latter being restricted to the case
$m=n$).

To begin we suppose $m=n$. In the case of $E((\lambda,\infty))$, by using the
limiting procedure discussed in the paragraph below (\ref{f3}), a formula can be
deduced from this 
for general $m \le n$. Our starting point is the formula \cite{SMM05}
\begin{equation}\label{14.1}
p(\lambda_1,\dots,\lambda_n) = \prod_{j=1}^n (r_j s_j)^n \prod_{j<k}^n
(\lambda_k - \lambda_j) {\cal I}(\{r_i\},\{s_i\},\{\lambda_i\})
\end{equation}
where, with the eigenvalues of $\Sigma_1^{-1}$, $\Sigma_2^{-1}$ given by
$\{r_i\}$, $\{s_i\}$ respectively,
\begin{eqnarray}\label{14.2}
&&  {\cal I}(\{r_i\},\{s_i\},\{\lambda_i\}) 
 \nonumber \\ && \qquad
= {1 \over n!}
\sum_{k_1 > k_2 > \cdots > k_n \ge 0}
\prod_{j=1}^n {(-1)^{k_j} \over k_j!} 
{\det [ r_j^{k_l}] \det [ \lambda_j^{k_l}] \det [s_j^{k_l}] \over
\prod_{j<l}^n (k_l - k_j) (r_l - r_j) (s_l - s_j) }.
\end{eqnarray}
In \cite[Lemma 5]{SMM05} it is proved that ${\cal I}$ is bounded by an exponentially
decaying function in $\lambda_j$ for each $j=1,\dots,n$.

Consider first $E((0,\lambda))$. We thus seek to integrate each $\lambda_i$ in (\ref{14.1})
over $(\lambda, \infty)$. For this we take inspiration from \cite{SMM05} and note from 
the Vandermonde determinant evaluation that 
\begin{equation}\label{15.1}
 \prod_{j<k}^n (\lambda_k - \lambda_j) =
\prod_{j=1}^n \Big ( {\lambda_j - \lambda \over \lambda^{1/2} } \Big )^{n-1}
\det \Big [ \Big ( {\lambda_j \over \lambda_j - \lambda} \Big )^{k-1} 
\Big ]_{j,k=1,\dots,n}.
\end{equation}
Expanding out the determinant according to its definition, substituting in (\ref{14.1})
and integrating gives
\begin{eqnarray}\label{16.1}
 E((0,\lambda)) \Big |_{m=n} & = & 
 \prod_{j=1}^n (r_j s_j)^n \,
\lambda^{-n(n-1)/2}  \sum_{P \in S_N} \varepsilon(P)
\int_\lambda^\infty d\lambda_1 \cdots \int_\lambda^\infty d\lambda_N \,
\nonumber \\
&& \times \prod_{j=1}^n (\lambda_j - \lambda)^{n - P(j)} \lambda_j^{P(j) - 1}
{\cal I}(\{r_i\},\{s_i\},\{\lambda_i\}).
\end{eqnarray}

For each integration variable, we integrate by parts $P(j)-1$ times, making use of the
simple formula
$$
(\lambda_j - \lambda)^m = {1 \over m + 1} {\partial \over \partial
\lambda_j} (\lambda_j - \lambda)^{m+1},
\qquad m \ne - 1.
$$
For $m > -1$, $(\lambda_j - \lambda)^{m+1}$ vanishes at $\lambda_j = \lambda$
(this is part of the motivation for the manipulation (\ref{15.1})), while from the
remark below (\ref{14.2}) the factor involving ${\cal I}$ vanishes at $\lambda_j =
\infty$. Hence in the integration by parts there is no contribution from the end points.
We must compute the partial derivatives with respect to $\lambda_j$ of
$\lambda_j^{P(j)-1} {\cal I}$. For this note that the only term dependent of $\lambda_j$ in
${\cal I}$ is $\det[\lambda_j^{k_l}]$, and
$$
\prod_{j=1}^n \lambda_j^{P(j)-1} \det [\lambda_j^{k_l}] =
\det [ \lambda_j^{k_l + P(j) - 1} ].
$$
Performing $P(j)-1$ integration by parts in each variable $\lambda_j$ thus gives
\begin{eqnarray}
 E((0,\lambda)) \Big |_{m=n} & = & 
\prod_{j=1}^n (r_j s_j)^n \,
{\lambda^{-n(n-1)/2} \over n! } \sum_{P \in S_N} \varepsilon(P)
\int_\lambda^\infty d\lambda_1 \cdots \int_\lambda^\infty d\lambda_N \,
\nonumber \\
&& \times \prod_{j=1}^n (\lambda_j - \lambda)^{(n - 1)}
\sum_{k_1 > \cdots > k_n \ge 0} \prod_{j=1}^n
{(-1)^{k_j} \over k_j!} \nonumber \\
&& \times {\det [ r_j^{k_l}] \det \Big [ 
\prod_{p=1}^{P(j)-1} \Big ( - {k_l + p \over n - p} \Big ) 
\lambda_j^{k_l} \Big ] \det [s_j^{k_l}] \over
\prod_{j<l}^n (k_l - k_j) (r_l - r_j) (s_l - s_j) }.
\end{eqnarray}

Next we want to integrate row by row in the determinant. Although the integrand decays
exponentially at infinity, this gives divergent integrals, as a result of interchanging
the order of summation and integration. To overcome this, write
$$
\int_\lambda^\infty d\lambda_j = \lim_{L \to \infty} \int_\lambda^L  d\lambda_j
\qquad (j=1,\dots,n),
$$
and so interchange only the finite range integrals with the summation of $\{h_j\}$.
We see the resulting one dimensional integrals are the same down each column of the
determinant and so can be factored. Furthermore, the sum over $P \in S_N$ then simply
interchanges rows in the determinant, which is compenstated for by $\varepsilon(P)$,
thus contributing an overall factor of $n!$. Hence
\begin{eqnarray}\label{18}
&& E((0,\lambda)) \Big |_{m=n} =   \prod_{j=1}^n (r_j s_j)^n
\, \lambda^{-n(n-1)/2}  \lim_{L \to \infty}
\sum_{k_1 > \cdots > k_n \ge 0} \prod_{j=1}^n
{(-1)^{k_j} \over k_j!}
 {\det [ r_j^{k_l}] 
\det [s_j^{k_l}] \over
\prod_{j<l}^n (k_l - k_j) (r_l - r_j) (s_l - s_j) } \nonumber
\\ 
&&\qquad \prod_{l=1}^n \int_\lambda^L (t - \lambda)^{n-1} t^{k_l} \, dt \,
\det \Big [ \prod_{p=1}^{j-1} \Big ( - {k_l + p \over n - p} \Big ) \Big ]_{j,l=1,\dots,n}.
\end{eqnarray}
As noted in \cite{SMM05}, it is straightforward to verify that
$$
\det \Big [ \prod_{p=1}^{j-1} \Big ( - {k_l + p \over n - p} \Big ) \Big ]_{j,l=1,\dots,n}
= (-1)^{n(n-1)/2}\prod_{j=1}^{n-1} {1 \over j^j} \prod_{j<l}^n (k_l - k_j),
$$
thus cancelling $\prod_{j<l}^n (k_l - k_j)$ and reducing (\ref{18}) to
\begin{eqnarray}\label{18.1}
E((0,\lambda))\Big |_{m=n} & = & 
 \prod_{j=1}^n (r_j s_j)^n \,
(-\lambda)^{-n(n-1)/2}  \prod_{j=1}^{n-1} {1 \over j^j} 
 \lim_{L \to \infty}
\sum_{k_1 > \cdots k_n \ge 0} \nonumber \\
&& \times \prod_{j=1}^n \Big ( \int_{\lambda}^L (t - \lambda)^{n-1}
{(-t)^{k_j} \over k_j!} \Big )
 {\det [ r_j^{k_l}] 
\det [s_j^{k_l}] \over
\prod_{j<l}^n (k_l - k_j) (r_l - r_j) (s_l - s_j) }
\end{eqnarray}

The lattice version of the well known Heine formula from random matrix theory
(see e.g.~\cite{Fo02}),
\begin{eqnarray*}
&&\int_I d \mu(x_1) \cdots \int_I d \mu(x_N) \,
\det [\phi_j(x_k) ]_{j,k=1,\dots,N}
\det [\psi_j(x_k) ]_{j,k=1,\dots,N} \nonumber \\
&& \qquad = N! \det [ \int_I \phi_j(x) \psi_k(x) \, d\mu(x)
]_{j,k=1,\dots,N}
\end{eqnarray*}
namely
$$
\sum_{k_1 > \cdots > k_n \ge 0} \det [a_i^{k_j}] \det [b_i^{k_j}] \prod_{i=1}^n
w(k_i) = \det \Big [ \sum_{p=0}^\infty w(p) (a_i b_j)^p \Big ]_{i,j=1,\dots,n}.
$$
referred to in \cite{SMM05} as the Cauchy-Binet formula, allows the sum in (\ref{18.1}) to be
computed. Taking then the limit $L \to \infty$ gives the sought determinant formula
\begin{eqnarray}\label{19.1}
E((0,\lambda))\Big |_{m=n} & = & 
 \prod_{j=1}^n (r_j s_j)^n \,
(-\lambda)^{-n(n-1)/2}  \prod_{j=1}^{n-1} {1 \over j^j}
{1 \over \prod_{j<l}^n  (r_l - r_j) (s_l - s_j) } \nonumber \\
&& \times \det \Big [ \int_{\lambda}^\infty (t - \lambda)^{n-1} e^{- t r_j s_l} \, dt
\Big ]_{j,l=1,\dots,n}.
\end{eqnarray}
And changing variables $t \mapsto t + \lambda$ in the integral allows (\ref{19.1}) to be
simplified further, giving
\begin{equation}\label{3.9}
E((0,\lambda))\Big |_{m=n} = \prod_{j=1}^{n-1} j! {1 \over \prod_{j<l}^n  (-\lambda)
(r_l - r_j) (s_l - s_j) }
\det [ e^{- \lambda r_j s_l} ]_{j,l=1,\dots,n}.
\end{equation}
Curiously this is the
Harish-Chandra/Itzykson-Zuber
matrix integral evaluation used in going from (\ref{5f}) to (\ref{6.1}) and so we have the
matrix integral representation 
\begin{equation}\label{3.10}
E((0,\lambda))\Big |_{m=n} = 
\int e^{- {\rm Tr}(\lambda R V^\dagger S V) } \, [V^\dagger dV],
\end{equation}
where $R,S$ are Hermitian matrices has eigenvalues $\{r_i \}$, $\{s_i\}$
respectively, and $[V^\dagger dV]$ denotes the normalized Haar measure,
$ \int [V^\dagger dV] = 1$.
With $m=n$, in the limit $s_1,\dots,s_n \to 1$ the p.d.f.~(\ref{ZZ1}) reduces to
(\ref{ZZ}). In keeping with this we can check that (\ref{3.9}) reduces to
(\ref{2.12}) (with $s_j \mapsto r_j$ $(j=1,\dots,n$) in the latter).

We turn our attention now to $E((\lambda, \infty)$. For this we use a minor rewrite of
(\ref{15.1})
$$
 \prod_{j<k}^n (\lambda_k - \lambda_j) =
\prod_{j=1}^n \Big ( {\lambda - \lambda_j \over \lambda^{1/2} } \Big )^{n-1}
\det \Big [ \Big ( {\lambda_j \over \lambda - \lambda_j} \Big )^{k-1} \Big ].
$$
in (\ref{14.1}) so that the analogue of (\ref{16.1}) reads
\begin{eqnarray*}
 E((\lambda,\infty)) \Big |_{m=n} & = &
 \prod_{j=1}^n (r_j s_j)^n \,
\lambda^{-n(n-1)/2}  \sum_{P \in S_N} \varepsilon(P)
\int_0^\lambda d\lambda_1 \cdots \int_0^\lambda d\lambda_N \,
\nonumber \\
&& \times \prod_{j=1}^n (\lambda - \lambda_j)^{n - P(j)} \lambda_j^{P(j) - 1}
{\cal I}(\{r_i\},\{s_i\},\{\lambda_i\}).
\end{eqnarray*}
The procedure of going from (\ref{16.1}) to (\ref{19.1}) can now be enacted.
(Note that in the integration by parts the factor
$$
\prod_{j=1}^n ( \lambda - \lambda_j)^{n - P(j)} \lambda_j^{P(j) - 1}
$$
ensures that the integrand vanishes at the end points.) We thus arrive at the
determinant evaluation
\begin{equation}\label{22}
 E((\lambda,\infty)) \Big |_{m=n} =
 \prod_{j=1}^{n-1} {1 \over j^j}
 {\prod_{j=1}^n (\lambda r_j s_j)^n \over \prod_{j < l}^n
(-\lambda) (r_l - r_j) (s_l - s_j) }
\det \Big [ \int_0^1 (1 - t)^{n-1} e^{- \lambda r_j s_l t} \, dt
\Big ]_{j,l = 1,\dots,n}.
\end{equation}

Analogous to the remark in the paragraph below (\ref{3.10}), we must have that for
$s_1,\dots,s_n \to 1$ (\ref{22}) coincides with (\ref{7.1a}) (after setting
$s_j \mapsto r_j$ $(j=1,\dots,n)$ in the latter). Now, taking the limit
$s_1,\dots, s_n \to 1$ in (\ref{22}) gives
\begin{equation}\label{sm}
\Big ( {1 \over (n-1)!} \Big )^{n-1}
{\prod_{j=1}^n (\lambda r_j)^n \over \prod_{j<l} (-\lambda) (r_l - r_j) }
\det \Big [ \int_0^1 e^{-\lambda r_j t} {d^{k-1} \over d t^{k-1} } (
t^{k-1} (1 - t)^{n-1} ) \, dt \Big ]_{j,k=1,\dots,n}
\end{equation}
Expanding the derivatives using the product rule, we see  after elementary column
operations that the determinant in (\ref{sm}) is equal to
\begin{eqnarray}\label{bb}
&& \det \Big [ \int_0^1 e^{- \lambda r_j t} t^{k-1} {d^{k-1} 
\over d t^{k-1} } (
t^{k-1} (1 - t)^{n-1} ) \, dt \Big ]_{j,k=1,\dots,n}
\nonumber \\
&& \qquad = \prod_{k=1}^{n-1} {(n-1)! \over (n-k)!} \,
\det \Big [ \int_0^1 e^{- \lambda r_j t}  (1 - t)^{n-1} \Big ( {t \over 1 - t}
\Big )^{k-1} \, dt \Big ]_{j,k=1,\dots,n}.
\end{eqnarray} 
But the determinant in (\ref{bb}) can be written
\begin{eqnarray}\label{aa}
&&\int_0^1 dt_1 \cdots \int_0^1 dt_n \, \prod_{j=1}^n e^{- \lambda r_j t_j} (1 - t_j)^{n-1}
\det \Big [  \Big ( {t_j \over 1 - t_j}
\Big )^{k-1}  \Big ]_{j,k=1,\dots,n} \nonumber \\
&& \quad
= \int_0^1 dt_1 \cdots \int_0^1 dt_n \, \prod_{j=1}^n e^{- \lambda r_j t_j}
\prod_{j<k}(t_k - t_j) 
\det \Big [ \int_0^1 e^{- \lambda r_j t} t^{k-1} \, dt \Big ]_{j,k=1,\dots,n}
\end{eqnarray}
Substituting (\ref{aa}) in (\ref{bb}), then substituting the result in
(\ref{sm}) reclaims (\ref{7.1a}) in the case $m=n$.

It remains to apply the limiting procedure discussed in the paragraph below 
(\ref{f3}) to deduce from (\ref{22}) the evaluation for general $m \le n$.
Noting the asymptotic expansion
$$
\int_0^1 (1 - t)^{n-1}  e^{- \lambda r_j s_l t} \, dt \: \sim \:
\sum_{p=1}^n (-1)^{p-1} {(n-1) \cdots (n - p + 1) \over ( \lambda r_j s_l)^p }
$$
the required limits can be taken to give
\begin{eqnarray}
 E((\lambda,\infty)) & = & (-1)^{(n-m)(n-m-1)/2}
 \prod_{j=1}^{n-1} {1 \over j^j} \prod_{p=1}^{n-m-1} {\Gamma(n) \over \Gamma(n-p) }
 {(\prod_{j=1}^m  r_j )^n (\prod_{j=1}^n \lambda  s_j)^n \over \prod_{j < l}^m
(r_l - r_j) \prod_{j<l}^n \lambda (s_l - s_j) } \nonumber \\
&& \qquad \times
\det \left [ \begin{array}{c}
[\int_0^1 (1 - t)^{n-1} e^{- \lambda r_j s_l t} \, dt]_{j=1,\dots,m \atop
l=1,\dots,n} \\
{}[(\lambda s_l)^{-j} ]_{j=1,\dots,n-m \atop l=1,\dots,n} \end{array} 
\right ]
\end{eqnarray}

\section*{Acknowledgement}
This work was supported by the Australian Research Council.


\begin{thebibliography}{10}

\bibitem{BBP05}
J.~Baik, G.~Ben Arous, and S.~P\'ech\'e.
\newblock Phase transition of the largest eigenvalue for nonnull complex sample
  covariance matrices.
\newblock {\em Annals of Prob.}, 33:1643--1697, 2005.

\bibitem{BK04}
P.M. Bleher and A.~Kuijlaars.
\newblock Integral representations for multiple {H}ermite and {L}aguerre
  polynomials.
\newblock arXiv:math.CA/0406616.

\bibitem{CWZ03}
M.~Chiani, M.Z. Win, and A.~Zanella.
\newblock On the capacity of spatially correlated {MIMO} {R}ayleigh-fading
  channels.
\newblock {\em IEEE Trans. Inform. Theory}, 49:2363--2371, 2003.

\bibitem{Ed88}
A.~Edelman.
\newblock Eigenvalues and condition numbers of random matrices.
\newblock {\em SIAM J. Matrix Anal. Appl.}, 9:543--560, 1988.

\bibitem{Fo02}
P.J. Forrester.
\newblock Log-gases and {Random} {Matrices}.
\newblock www.ms.unimelb.edu.au/\~{}matpjf/matpjf.html.

\bibitem{Fo93c}
P.J. Forrester.
\newblock Exact results and universal asymptotics in the {Laguerre} random
  matrix ensemble.
\newblock {\em J. Math. Phys.}, 35:2539--2551, 1993.

\bibitem{FH94}
P.J. Forrester and T.D. Hughes.
\newblock Complex {Wishart} matrices and conductance in mesoscopic systems:
  exact results.
\newblock {\em J. Math. Phys.}, 35:6736--6747, 1994.

\bibitem{GS00}
H.~Gao and P.J. Smith.
\newblock A determinant representation for the distribution of quadratic forms
  in complex normal vectors.
\newblock {\em J. Mult. Anal.}, 73:155--165, 2000.

\bibitem{GN99}
A.K. Gupta and D.K. Nagar.
\newblock {\em Matrix variate distributions}.
\newblock Chapman \& Hall/CRC, Boca Raton, 1999.

\bibitem{Ja64}
A.T. James.
\newblock Distributions of matrix variate and latent roots derived from normal
  samples.
\newblock {\em Ann. Math. Statist.}, 35:475--501, 1964.

\bibitem{Jo01}
I.M. Johnstone.
\newblock On the distribution of the largest principal component.
\newblock {\em Ann. Math. Stat.}, 29:295--327, 2001.

\bibitem{KE04}
P.~Koev and A.~Edelman.
\newblock The efficient evaluation of the hypergeometric function of a matrix
  argument.
\newblock arXiv:math.PR/0505344, 2005.

\bibitem{Mu82}
R.J. Muirhead.
\newblock {\em Aspects of multivariable statistical theory}.
\newblock Wiley, New York, 1982.

\bibitem{Or04}
A.Y. Orlov.
\newblock New solvable matrix integrals.
\newblock {\em Int. J. Mod. Phys. A}, 19:276--293, 2004.

\bibitem{RVA05}
T.~Ratnarajah, R.~Vaillancourt, and M.~Alvo.
\newblock Eigenvalues and condition numbers of complex random matrices.
\newblock {\em SIAM J. Matrix Anal. Appl.}, 26:441--456, 2005.

\bibitem{SM04}
S.H. Simon and A.~L. Moustakas.
\newblock Eigenvalue density of correlated random {W}ishart matrices.
\newblock {\em Phys. Rev. E}, 69:065101(R), 2004.

\bibitem{SMM05}
S.H. Simon, A.~L. Moustakas, and L.~Marinelli.
\newblock Capacity and character expansions: moment generating function and
  other exact results for {MIMO} correlated channels.
\newblock arXiv:cs.IT/0509080, 2005.

\bibitem{TV04}
A.M. Tulino and S.~Verd\'u.
\newblock Random matrix theory and wireless communications.
\newblock volume~1 of {\em Foundations and {T}rends in {C}ommuncations and
  {I}nformation {T}heory}, pages 1--182. Now, 2004.

\end{thebibliography}

\end{document}